\documentclass[12pt]{article}

\usepackage{amsmath}
\usepackage{amssymb}
\usepackage{verbatim}

\makeatletter

   \@addtoreset{equation}{section}
\makeatother

\begin{document}

\vskip 12mm

\begin{center} 
{\Large \bf  Irregular vertex operators for  irregular conformal blocks}
\vskip 10mm
{ \large  Dmitri Polyakov$^{a,b,}$\footnote{email:polyakov@scu.edu.cn;polyakov@sogang.ac.kr}
and  Chaiho Rim$^{c,}$\footnote{email:rimpine@sogang.ac.kr} }

\vskip 8mm
$^{a}$ {\it  Center for Theoretical Physics, College of Physical Science and Technology}\\
{\it  Sichuan University, Chengdu 6100064, China}\\
\vskip 2mm

$^{b}$ {\it Institute of Information Transmission Problems (IITP)}\\
{\it  Bolshoi Karetny per. 19/1, Moscow 127994, Russia}\\
\vskip 2mm
$^{c}$ {\it Department of Physics and Center for Quantum Spacetime}\\
{\it Sogang University, Seoul 121-742, Korea}

\end{center}

\vskip 15mm

\begin{abstract}
We construct the free field representation of  irregular vertex operators of arbitrary rank 
which generates simultaneous eigenstates of positive modes of Virasoro and W symmetry generators. 
The irregular vertex operators turn out to be the exponentials of combinations of derivatives of
Liouville or Toda  fields, creating irregular coherent states. We compute
examples of correlation functions of these operators and study their operator algebra.

\end{abstract}
%\vfill
%\end{titlepage}

\vskip 12mm

\setcounter{footnote}{0}

\section{Introduction} 

Primary vertex operators in two-dimensional conformal field theory
are the objects playing  a crucial role in the AGT conjecture \cite{AGT},
connecting regular Liouville conformal blocks to Nekrasov's partition function
\cite{Nekrasov}
on the Coulomb branch of $N=2$
supersymmetric gauge theories in four dimensions.
Among the interplay between the four dimensional  gauge theory and 
two dimensional CFT, there appears a non-trivial IR fixed point,
Argyres-Douglas type theory \cite{AD, APSW}.
This class of theories does not allow marginal deformations
and is  described in terms of colliding limit of the primary 
vertex operators.  The operator of rank $q$, obtained from the colliding limit
\cite{EM_2009,GT_2012},
generates a irregular state of rank $q$ 
when applied to the vacuum. 
The irregular state is annihilated by $L_k$ with $k > 2q$
but becomes 
 a simultaneous eigen-state of positive Virasoro generators
$L_k$ with $q \le k \le 2q$.
This irregular state is called Gaiotto state \cite{G_2009}
or Whittaker state \cite{Whittaker}.  
The usual regular primary state corresponds to the rank 0 state 
 
One obvious try to construct the 
irregular vertex operators (IVO) or the irregular conformal states
was the construction of  the state as the combination of the  primary state 
and its descendents \cite{G_2009,Whittaker,MMM_0909,KMST_2013,Bonelli_2011}. 
However, the attempt to find the irregular state beyond the rank 1 has met a serious difficulty to fix the coefficient 
if one uses the 
fact that the state is the simultaneous eigenstate of the positive Virasoro generators only. 
The state thus constructed  has undetermined parameters which should be further fixed by the 
consistency condition with the lower mode $L_{k<q} $ \cite{CRZ_2014}.   

In this paper we re-consider the irregular vertex operator
directly in terms of free bosonic field representation. 
To get an idea, we notes that the two-point conformal block,
one primary vertex operator at infinity and one irregular vertex operator 
at the origin, is given as the irregular matrix model (IMM) \cite{EM_2009,NR_2012}
which has the form of Penner-type matrix models
\begin{eqnarray}
Z=\int\prod_{i=1}^Nd\lambda_i\prod_{1\leq{j}<k\leq{N}}
(\lambda_j-\lambda_k)^{-2b^2}e^{-2b\sum_{i}V(\lambda_i)}
\end{eqnarray}
where the potential has the logarithmic term together with the  inverse power-like contributions
\begin{eqnarray}
V(\lambda_i)=c_0 {\log(\lambda_i)}
-\sum_{j=1}^{q}{{c_{j}}\over{j(\lambda_i)^j}} \,.
\end{eqnarray}
%IMM obeys the neutrality condition 
%\begin{eqnarray}
%c_0+\alpha_\infty+bN=Q_b = (b + \frac 1b){\sqrt{2}}
%\end{eqnarray} 
%where $\alpha_\infty$ is the Liouville charge at infinity. 
%The eigenvalue $\Lambda_k$  of the  positive Virasoro generator  $L_k$ 
%on the irregular conformal state is related with the coefficients $c_j$'s 
%is related as $\Lambda_k = (k+1) Q_b c_k - \sum_{p \ge 0} c_p c_{k-p}$. 
Seiberg-Witten curve obtained from the loop equation of IMM
has the quadratic form and IMM is  expected to reproduce the 
instanton contributions to the partition functions in the Argyres-Douglas
theories according to AGT.  
The irregular conformal blocks  (ICB)  are in general not simple objects to explore,
even though the IMM approach to the ICB 
provides a relatively simple procedure but needs tedious steps to find ICB 
working with  loop equations.
Therefore, it is desirable to find IVO directly from the eigenvalue constraints 
using the (Liouville) free fields and provide  ICB in terms of IVO directly.  

The general feature of the potential term of IMM
is that IVO can be represented in terms of modified vertex operators 
which contains finite number of derivatives of the  Liouville fields \cite{CRZ_2015}.
However, it is yet to be  checked if the modified primary operator 
indeed represents the irregular vertex operator. 
In this paper we construct the free field representation of  IVO explicitly 
without resorting to the ICB  or IMM  
but only using the fact that IVO produces the simultaneous eigenstates of 
positive  generators. 
For the Virasoro IVO, one has the conditions:
\begin{align}
\left[  L_k, I_q \right]  =\rho_k I_q ~~ (q\leq{k}\leq{2q}); 
~~~~
%\nonumber \\
\left[ L_k,  I_q \right] =0~~ (k> 2q)
\label{eq:Virasoro-constraint}
\end{align}
where $ I_q $ is the IVO of rank $q$ and $ \rho_k $ is the eigenvalue 
of the positive mode Virasoro generator $ L_k$. 

In case of two or more copies of the Liouville fields as in the 
Toda field theories, IVO  can have more constraints 
to incoorporate the higher spin symmetry
in addition to \eqref{eq:Virasoro-constraint}. 
For example, with two fields, IVO subjects to the $W^{(3)}$  symmetry constraints:
\begin{align}
[ W^{(3)}_k, I_q]  =\omega^{(3)}_k I_q ~~(2q\leq{k}\leq{3q}); ~~~~
%\nonumber \\
[W^{(3)}_k, I_q] = 0 ~~(k> 3q )
\end{align}
where $W^{(3)}_k$ is the $k$-th  mode of the spin 3
$W$-current and  $\omega^{(3)}_k$  is  its  eigenvalue. 
The corresponding irregular matrix models  can be obtained 
from the  colliding limit of the  $A_2$ Toda field  theory,
whose loop equation provides the cubic form of the Seiberg-Witten curve
and flow equations corresponding to  $W^{(3)}$ symmetry 
\cite{CRZ_2015,CR_2015}. 
It is  generally expected that IMM obtained from the colliding limit 
of $A_r$ Toda field theory results in the Seiberg-Witten curve with the $(r+1)$-th power term
and flow equations of  $W^{(r+1)}$ symmetry. 
The corresponding IVO can be determined 
by the generalized constraints due to  $W^{(r+1)}$ symmetry:
\begin{align*}
[ W^{(r+1)}_k, I_q]  =\omega^{(3)}_k I_q ~~(rq\leq{k}\leq{(r+1)q}); ~~~~
%\nonumber \\
[W^{(r+1)}_k, I_q] = 0 ~~(k> (r+1)q )
\end{align*}

This paper is organized as follows.
In section 2,  we consider the case with one free bosonic field 
which has Virasoro symmetry. We first develop the free field representation 
of the Virasoro IVO of rank 1 by solving the Virasoro constraint, reproducing the deformed
Penner-type  potential of the matrix model approach. 
We then extend this construction to higher ranks and present 
the general structure of IVO of arbitrary ranks.
The explicit coordinate dependence of ICB constructed from N-point  IVO correlator 
is given in free field formalism.

In section 3, we extend this construction to the system of two bosonic fields
so that IVO obeys the $W^{(3)}$ symmetry.
We explicitly check that IVO of lower rank has the similar free field representation 
as in the Virasoro case. 
The eigenvalues fix IVO with algebraic polynomial equations. 
 
Section 4 is the conclusion where IVO of arbitrary rank $q$ with $W^{(r+1)}$-symmetry
is given and its eigenvalues are presented explicitly in terms of the coefficients of IVO
for $W^{(3)}$ case.
In addtion, some of  physical implications of IVO are speculated. 

 %=========================================%
\section{Irregular vertex operator with Virasoro symmetry} 
 %=========================================%

In this section we demonstrate the explicit construction for IVO in terms of one free bosonic field.
Before we demonstrate the explicit ansatz,
it is useful to comment on the structure of the answer that we expect
and its relation to the colliding limit.

The irregular blocks of rank $q$ essentially emerge as a result of the normal
ordering $q+1$ Liouville vertex operators colliding at the same point.
Let us consider the example of two vertex operators first.
The operator product between two exponential operators at
points $z_1$ and $z_2$ around $z_2$ is given by 
\begin{eqnarray}
e^{\alpha\phi}(z_1)e^{\beta\phi}(z_2)=(z_{12})^{-\alpha\beta}\sum_{n=0}^\infty
(z_{12})^n:B^{(n)}_\alpha(\phi)e^{(\alpha+\beta)\phi}:(w)
\end{eqnarray}
where $z_{12}= z_1-z_2$.  
 $B^{(n)}_\alpha$ are the normalized Bell polynomial of the derivatives
of $\phi$ and are defined as \cite{polyakov_2015}
\begin{eqnarray}
B^{(n)}_\alpha=\sum_{p=1}^n\alpha^p\sum_{n|k_1...k_p}
{{\partial^{k_1}\phi...\partial^{k_p}\phi}\over{k_1!q_{k_1}!...k_p!q_{k_p}!}}\,.
\end{eqnarray}
Here the sum is taken over the ordered length $p$ partitions
of $n$ $(1\leq{p}\leq{n})$: $n=k_1+...+k_p$;
$k_1{\leq}k_2...\leq{k_p}$
and $q_{k_j}$ is the multiplicity of an element $k_j$ in the partition.
The operator product 
for three  operators at $z_1,z_2,z_3$ colliding at $z_1$
is similarly given by
\begin{align}
e^{\alpha\phi}(z_1)e^{\beta\phi}(z_2)e^{\gamma\phi}(z_3)
=&(z_2-z_1)^{-\alpha\beta}(z_3-z_1)^{-\gamma(\alpha+\beta)}
\sum_{n_2=0}^{\infty}\sum_{n_1=0}^{\infty}\sum_{k=0}^{n_1}
(z_2-z_1)^{n_1}(z_3-z_1)^{n_2-k}
\nonumber \\
\times
&~~~
{{\Gamma(-\beta\gamma+1)}\over{k!\Gamma(-\beta\gamma+1-k)}}
:{B^{(n_1-k)}_\beta}{B^{(n_2)}_\gamma}e^{\alpha+\beta+\gamma}:(z_1)\,.
\end{align}
One may have in general,
for N vertices at $z_1,...z_N$ around $z_1$  
\begin{align}
&e^{\alpha_1\phi}(z_1)...e^{\alpha_N}(z_N)
=\prod_{p=2}^N(z_{k1})^{-\alpha_p(\alpha_1+...\alpha_{p-1})}
\sum_{n_1,...,n_{N-1}}\sum_{k_1,...k_{N-2}}\sum_{q_1....q_{N-2}}
(z_{21})^{n_1}(z_{31})^{n_2-k_1}...
\nonumber \\
&~~~~~~~~~~~
\times
(z_{N1})^{n_{N-1}-k_{N-2}}
%\nonumber \\
\prod_{j=1}^{N-1}
\lambda_{\lbrace{n,k,q}\rbrace}
:B^{(n_j-q_j(k_1,...k_{N-1}))}_{\alpha_{j+1}}
e^{(\alpha_1+...+\alpha_N)\phi}:(z_1)
\label{eq:N-fusion}
\end{align}
with 
the $q$-numbers satisfying
\begin{eqnarray}
\sum_{j=1}^{N-2}k_j=\sum_{j=1}^{N-2}q_j
\end{eqnarray}
and $\lambda_{\lbrace{n,k,q}\rbrace}$ are some constants which are 
straightforward to evaluate but whose explicit form is of no importance to us.
IVO is then obtained by taking 
the operator product \eqref{eq:N-fusion} inside correlators  and taking the simultaneous 
limits $z_j\rightarrow{z_1};j=2,...,N$
maintaining $\sum \alpha_i z_i^k$ finite for $k=0,1,\cdots, n$. 
All the operators appearing on the right hand side of the operator product
 \eqref{eq:N-fusion}  shall appear in the expression for IVO; thus, IVO of
any rank $q$ must contain infinite number of terms, having the general form
\begin{equation}
I_q\sim\sum_{n_1,...,n_q=0}^\infty\lambda_{n_1...n_q}
{e^{\sum_{j=1}^{q+1}\alpha_j\phi}}B_{\alpha_1}^{(n_1)}...B_{\alpha_{q}}^{(n_{q})}
\end{equation}
where the $\lambda$-coefficients must be determined from the Virasoro constraints
\eqref{eq:Virasoro-constraint}. 
It is noteworthy that the objects similar to that type appear in string 
field theory as analytic solutions of the equations of motion,
presumably describing the collective higher spin vacuum state \cite{polyakov_2015}.

However, taking the colliding limits directly for the expansion in \eqref{eq:N-fusion} 
is obviously a tedious procedure and technically seems to be beyond
control in an arbitrary case. 
In addition, the products of Bell polynomial operators, generally lead to tedious
recursion relations, and are hard to solve analytically. 
Therefore, it shall be better to work in a different operator basis, 
namely, ifield derivative basis 
so that we can apply the Virasoro constraints \eqref{eq:Virasoro-constraint}
directly to the operator basis.  

To start, we look for the solution IVO of rank one in the form
\begin{eqnarray}
I_1=\sum_{N_1,N_2=0}^\infty
\lambda_{N_1N_2}\phi^{N_1}(\partial\phi)^{N_2}
\end{eqnarray} 
Second and higher derivatives are not allowed
since  $I_1$ should be annihilated  
by all the $L_{k}$-generators  with $k>2$: 
Note that $L_k=\oint{{dz}\over{2i\pi}}z^{k+1}T(z)$
and the stress-energy tensor with background charge $Q$
\begin{eqnarray}
T(z)=-{1\over{2}}(\partial\phi)^2+{Q\over{2}}\partial^2\phi
\end{eqnarray}
will have the leading OPE singularity of the order $\sim{(z_1-z_2)^{-3}}$
with $I_1$
since we are using the free field normalization 
$\langle \phi(z) \phi(w) \rangle= - \log (z-w)$. 

The solution is obtained if one finds the  generating function $F(x,y)$ 
of two variables with the same  $\lambda_{N_1N_2} $: 
\begin{eqnarray}
F(x,y)=\sum_{N_1,N_2}\lambda_{N_1N_2}x^{N_1}y^{N_2}
\end{eqnarray}
The eigenvalue constraint 
\begin{align}
\left [  L_2,  I _1  \right] =\rho_2 I_1  
\end{align}
leads to the relation 
\begin{eqnarray}
\rho_{2}\sum_{N_1,N_2=0}^\infty
\lambda_{N_1N_2}\phi^{N_1}(\partial\phi)^{N_2}
=
-\sum_{N_1=0,N_2=2}^\infty{N_2(N_2-1)}\lambda_{N_1N_2}\phi^{N_1}(\partial\phi)^{N_2-2}
\end{eqnarray}
It is easy to see that this equation is equivalent to a simple
partial differential equation on the generating function $F(x,y)$:

\begin{equation}
\partial^2_yF(x,y)=-\rho_2{F(x,y)}
\end{equation}
whose general solution is
\begin{equation}
F(x,y)=e^{i{\sqrt{\rho_2}}y}f(x)
\end{equation}
Similarly, the second eigenvalue problem:
\begin{equation}
[L_1, I_1] =\rho_1 I_1
\end{equation}
leads to the second recursion relation for $\lambda$:
\begin{eqnarray}
\rho_{1}\sum_{N_1,N_2=0}^\infty
\lambda_{N_1N_2}\phi^{N_1}(\partial\phi)^{N_2}
=
-\sum_{N_1=1,N_2=1}^\infty{N_1}N_2\lambda_{N_1N_2}\phi^{N_1-1}(\partial\phi)^{N_2-1}
\nonumber \\
+Q\sum_{N_1=1,N_2=1}^\infty{N_2}\lambda_{N_1N_2}\phi^{N_1}(\partial\phi)^{N_2-1}
\end{eqnarray}
leading to the second order differential equation
on $F$:
\begin{eqnarray}
-\partial_x\partial_y{F}+Q\partial_y{F}=\rho_1{F}
\end{eqnarray}
Substituting the general solution of the first equation
and identifying $f(x)$ we find the generating function to be given by
\begin{equation}
F(x,y)=e^{(i{{\rho_1}\over{\sqrt{\rho_2}}}+Q)x+i{\sqrt{\rho_2}}y}
\end{equation}
Accordingly, substituting for $\lambda_{N_1N_2}$ we find
that the expression for IVO of the rank 1  
\begin{equation}
I_1=:e^{(i{{\rho_1}\over{\sqrt{\rho_2}}}+Q)\phi+i{\sqrt{\rho_2}}\partial\phi}:
\end{equation}

The contribution to the Penner type potential
is given by the log of the leading order OPE term of $I_1(z_1)$ with 
a regular vertex $e^{\alpha\phi}(z_2)$. 
Expanding $I_1$ in terms of $\phi$ and $\partial\phi$ and exponentiang
one easily finds
\begin{equation}
V(z_{12})\sim
{(i{{\rho_1}\over{\sqrt{\rho_2}}}+Q) \log(z_{12})
+i{{\sqrt{\rho_2}}\over{z_{12}}}}
\end{equation}
reproducing the known result from the matrix model approach,
which leads to the identification of the potential coefficients with the 
eigenvalues. 

Next, let us consider the rank 2 case.
The eigenvalue constraints for the rank 2 are 
\begin{align}
[ L_k, I_2] &=\rho_k I_2; ~~(k=2,3,4) 
\nonumber \\ 
[ L_k, I_2] &=0; ~~~(k>4)\,.
\end{align}  
Accordingly,  the ansatz for the rank 2 will be 
\begin{equation}
I_2=\sum_{N_1,N_2,N_3}\lambda_{N_1N_2N_3}
\phi^{N_1}(\partial\phi)^{N_2}(\partial^2\phi)^{N_3}
\end{equation}
since $I_2$ is by construction annihilated by all $L_k$
for $k\geq{5}$. 
With the generating function 
$$F(x,y,z)=\sum_{N_1,N_2,N_3}\lambda_{N_1N_2N_3}x^{N_1}y^{N_2}z^{N_3} \,.$$
the eigenvalue constraints for $k=2,3$ and $4$
lead, in turn to the
characterictic 3 PDE's
\begin{eqnarray}
-4\partial^2_z{F}=\rho_4{F}
\nonumber \\
-2\partial_y\partial_z{F}=\rho_3{F}
\nonumber \\
-2\partial_x\partial_zF-\partial^2_yF+{3\over2}\partial_z{F}=\rho_2F\,.
\end{eqnarray}
The solution of the system is
\begin{eqnarray}
F=
e^{{i\over{{\sqrt{\rho_4}}}}((\rho_2-{{\rho_3^2}\over{\rho_4}})+{3\over2}Q)x
+i{{\rho_3}\over{{\sqrt{\rho_4}}}}y+{i\over2}{\sqrt{\rho_4}}z} \,,
\end{eqnarray}
leading to IVO:
\begin{eqnarray}
I_2=
e^{{i\over{{\sqrt{\rho_4}}}}((\rho_2-{{\rho_3^2}\over{\rho_4}})+{3\over2}Q)\phi
+i{{\rho_3}\over{{\sqrt{\rho_4}}}}\partial\phi+{i\over2}{\sqrt{\rho_4}}
\partial^2\phi}
\end{eqnarray}
with the corresponding contributions to the Penner's potential:
\begin{eqnarray}
V(z_{12})\sim
{i\over{{\sqrt{\rho_4}}}}((\rho_2-{{\rho_3^2}\over{\rho_4}})+{3\over2}Q) \log(z_{12})
+i{{\rho_3}\over{{\sqrt{\rho_4}}}}{z_{12}}^{-1}+{i\over2}{\sqrt{\rho_4}}
{z_{12}}^{-2}
\end{eqnarray}
It is not difficult to extend the same pattern to the higher ranks.
For the rank 3, the IVO ansatz will include the third derivatives of 
the Liouville field:
\begin{eqnarray}
I_2=\sum_{N_1,N_2,N_3,N_4}\lambda_{N_1N_2N_3N_4}
\phi^{N_1}(\partial\phi)^{N_2}(\partial^2\phi)^{N_3}(\partial^3\phi)^{N_4}\,.
\end{eqnarray}
The generating function will have 4 variables and satisfy
the system of 4 linear second order differential equations.
The computation similar to the above gives the answer for the
rank 3 irregular block in terms of the irregular vertex operator:
\begin{eqnarray}
I_3=e^{{{i}\over{{\sqrt{\rho_6}}}}\lbrace
(\rho_3-{{\rho_4\rho_5}\over{\rho_6}}+{{\rho_5^3}\over{\rho_6^2}}-2i
Q{\sqrt{\rho_6}})\phi
+(\rho_4-{{\rho_5^2}\over{\rho_6}})\partial\phi+{{\rho_5}\over2}\partial^2\phi
+{{\rho_6}\over{6}}\partial^3\phi\rbrace}
\end{eqnarray}
with the related contribution to the Penner's potential
\begin{eqnarray}
V_3(z_{12})\sim
{{i}\over{{\sqrt{\rho_6}}}}\lbrace
(\rho_3-{{\rho_4\rho_5}\over{\rho_6}}+{{\rho_5^3}\over{\rho_6^2}}-2i
Q{\sqrt{\rho_6}}) \log(z_{12})
\nonumber \\
+(\rho_4-{{\rho_5^2}\over{\rho_6}}){{z_{12}}^{-1}}+{{\rho_5}\over2}
{{z_{12}}^{-2}}
+{{\rho_6}\over{6}}{{z_{12}}^{-3}}\rbrace
\end{eqnarray}

It is now not difficult to guess the general structure 
of the answer for an arbitrary rank $q$:
IVO of the rank $q$  is given by 
\begin{equation}
I_q=:e^{ %{ i\over{{\sqrt{\rho_{2q}}}}}
\sum_{k=0}^{q}\alpha_k\partial^{k}\phi}:
\label{eq:IVO_q}
\end{equation}
with the related Penner type potential contribution
\begin{equation}
V_q(z_{12})\sim % {i\over{{\sqrt{\rho_{2q}}}}} \lbrace
\alpha_0 \log(z_{12})+\sum_{k=1}^q\alpha_q(z_{12})^{-k}
%\rbrace
\end{equation}
where
\begin{align}
- i \sqrt{\rho_{2q}}~
\alpha_{k}
&=
{{\rho_{q+k}}\over{k!}}
+\sum_{m=1}^{q-k-1}{{(-1)^m}\over{\rho_{2q}^m}}
\sum_{j=1}\sum_{(2m+1)q+k|q_1...q_j}n_{q_1...q_j}\rho_{q_1}...\rho_{q_j}
%\nonumber \\
-{{i{\sqrt{\rho_{2q}}}(q+1)Q}\over{2}}\delta_0^k
\nonumber \\
- i \sqrt{\rho_{2q}}~
\alpha_{q} &={{\rho_{2q}}\over{q!}}
\end{align}
with $ 0 \le k \le q-1$. 
$n_{q_1...q_j}$ are positive integers
and the second sum in the expression
$\sum_{(2m+1)q+k|q_1...q_j}$ is taken over all
possible length $j$ ordered partitions
of $$(2m+1)q+k=q_1+...+q_j$$
such that
$q\leq{q_1}\leq...\leq{q_j}\leq{2q}$ with the subsequent summation over the 
lengths.
Irregular conformal state obtained by 
IVO of the form \eqref{eq:IVO_q}  has the simultaneous eigenvalues 
$\rho_k $ of $L_k (k=q,...,2q)$ 
whose relation with $\alpha$-coefficients are given in terms 
of $q+1$ algebraic equations. 
The objects of the type (2.29) were also considered in
\cite{nagoya} in a  different context, as well as in 
\cite{lefloch}.

In addition, IVO of the form \eqref{eq:IVO_q} 
provides  the $N$-point ICB:
\begin{eqnarray}
<I_{q_1}(z_1)...I_{q_N}(z_N)>=
<\prod_{l=1}^N{:e^{\sum_{k_l=0}^{q_l}\alpha^{(q_l)}_{k_l}\partial^{k_l}\phi}(z_l):}>
\end{eqnarray}
where  $q_l(l=1,...,N)$ are the ranks of IVO.
% and 
%$\alpha_{k_l}^{(q_l)};k_l\leq{q_l}$  are the coefficients in front
%of $\partial^{k_l}\phi$ in the exponent of irregular vertex operator
%for the irregular block of rank $q_l$, defining the eigenvalues of
%$I_{q_l}$.
%
Below we shall compute this correlator in the limit
of zero Liouville cosmological constant, i.e. in the free field limit.
This calculation still holds at nonzero constant, as long
as  long as the neutrality condition
$\sum_{1 \le \ell \le N} \alpha^{(q_l)}_0  =  Q$ holds.  
Despite that, the free field calculation still makes sense even when 
the neutrality condition is not satisfied, since the irregular blocks are the objects
essentially appearing in the process of perturbative expansion 
in the screening operator, and, as such, get inserted inside
the free field correlators.

To compute the holomorphic correlator,
consider the functional integral
\begin{eqnarray}
<I_{q_1}(z_1)...I_{q_N}(z_N)>=
Z^{-1}\int{D\phi}
\prod_{l=1}^N{e^{\sum_{k_l=0}^{q_l}\alpha^{(q_l)}_{k_l}\partial^{k_l}\phi}(z_l)}
e^{-{1\over{8 \pi}}\int{d^2}z\partial\phi{\bar\partial\phi}}
\end{eqnarray}
This integral can be written
as
\begin{eqnarray}
<I_{q_1}(z_1)...I_{q_N}(z_N)> 
=\int{D\phi}e^{\int{d^2z}{1\over{8 \pi}}\partial\phi
{\bar\partial}\phi+\phi\sum_{l=1}^N\sum_{k_l=0}^{q_l}\alpha^{(q_l)}_{k_l}(-1)^{k_l}
\partial_z^{k_l}\delta^{(2)}(z-z_l)}
\end{eqnarray}
where we took the exponential insertions at $z_l$ inside the
$z$-integral by using the $\delta$-functions $\delta^{(2)}(z-z_l)$
and integrated $k_l$ times by parts for each derivative field 
$\partial^{k_l}\phi$.
This integral is now the Gaussian integral with the 
linear source term
\begin{eqnarray}
j(z,{\bar{z}})=\sum_{l=1}^N\sum_{k_l=0}^{q_l}\alpha^{(q_l)}_{k_l}(-1)^{k_l}
\partial_z^{k_l}\delta^{(2)}(z-z_l)
\end{eqnarray}
and its value simply equals to that of the generating functional
\begin{eqnarray}
W(j)=e^{\int{d^2z}\int{d^2w}j(z,{\bar{z}})j(w,{\bar{w}})
G(|z-w|)}
\end{eqnarray}
with $G(|z-w|)=-\log|z-w|^2$
Substituting for $j(z,{\bar{z}})$, integrating again by parts for each
term in the sum to bring the derivatives of the delta-dunctions
into the delta-functions - and finally integrating out the delta-functions,
we obtain:
\begin{align}
&<I_{q_1}(z_1)...I_{q_N}(z_N)>
\nonumber \\&~~~
= 
 e^{{1\over2}\sum_{l_1=1}^N\sum_{l_2=1}^N\sum_{k_{l_1}=0}^{q_{l_1}}
\sum_{k_{l_2}=0}^{q_{l_2}}\alpha^{(q_{l_1})}_{k_{l_1}}\alpha^{(q_{l_2})}_{k_{l_2}}
\int{d^2z}\int{d^2w}
\delta^{(2)}(z-z_{l_1})\delta^{(2)}(w-z_{l_2})\partial_z^{k_{l_1}}
\partial_w^{k_{l_2}}G(|z-w|)}
 \nonumber \\ 
&~~~
=
\prod_{l_1,l_2=1;l_1\neq{l_2}}
(z_{l_1}-z_{l_2})^{-\alpha^{q_{(l_1)}}_0\alpha^{q_{(l_2)}}_0}
% \nonumber \\ \times
e^{{1\over2}\sum_{l_1=1}^N\sum_{l_2=1;l_1\neq{l_2}}^N\sum_{k_{l_1}=0}^{q_{l_1}}
\sum_{k_{l_2}=0}^{q_{l_2}}{{(-1)^{k_{l_1}}(k_{l_1}+k_{l_2}-1)!
\alpha^{(q_{l_1})}_{k_{l_1}}\alpha^{(q_{l_2})}_{k_{l_2}}}\over{(z_{l_1}-z_{l_2})^{k_{l_1}+k_{l_2}}
}}
}\,.
\end{align}

This is the general answer.
For example, applied to the three-point function of the rank 2 blocks,
it gives:
\begin{eqnarray}
<I_2(z_1)I_2(z_2)I_2(z_3)>
={\lbrack}(z_1-z_2)(z_1-z_3)(z_2-z_3)
{\rbrack}^{{1\over{\rho_4}}(\rho_2-{{\rho_3^2}\over{\rho_4}}+{Q\over2})}
\nonumber \\
\times
e^{{{\rho_3^2}\over{\rho_4}}({1\over{(z_1-z_2)^2}}+
{1\over{(z_1-z_3)^2}}+{1\over{(z_2-z_3)^2}})+6\rho_4({1\over{(z_1-z_2)^4}}
+{1\over{(z_1-z_3)^4}}+{1\over{(z_1-z_2)^4}})
}\end{eqnarray}
where $\rho_{2,3,4}$ are the eigenvalues of $L_{2,3,4}$ respectively.
Note that the exponent only contains the $even$ powers of the inverse
$z_{ij}$, as it should be (otherwise the answer would have been
unphysical since for close $z_i$ and $z_j$ interchanging points
would e.g. make an infinitely large exponent out of infinitely small).

%\begin{center}
\section{Toda Generalizations and $W_n$ symmetries for Irregular Conformal Blocks}
%\end{center}

There was an insightful observation made in \cite{KMST_2013, CRZ_2015, CR_2015} that, 
when the Liouville theory is  exdended to $A_2$ Toda model containing two copies of the scalar field, 
the irregular conformal block of such a model possesses  additional
symmetries related to $W^{(3)}$ algebra.
Therefore, irregular state will be the eigenstgate not only 
of Virasoro generators $L_n$ with $q{\leq}{n}{\leq}{2q}$ 
but also of eigenvalues of the $W_n^{(3)}$ generators with 
$2q{\leq}n{\leq}3q$. 
This property has been demonstrated explicitly in
the random matrix model approach.  However,  
its generalization to  higher $W^{(n)}$ symmetry remain somewhat uncontrollable, 
As we shall demonstrate below, the whole construction
and its generalizations become much more simple and transparent in the vertex operator
formalism using the manifest free-field representation.
The free field representation for irregular rank $q$
conformal blocks involving $r$ scalar fields, leading to 
 irregular vertex operators of the type \eqref{eq:IVO_q},
 is given by
\begin{eqnarray}
I_{q|r}=
\sum_{N^{(1)}_1...N^{(r)}_q=0}^\infty
{\lambda_{N^{(1)}_1...N^{(1)}_q|N^{(2)}_1...N^{(2)}_q|
...|N^{(r)}_1...N^{(r)}_q}}
(\phi^{(1)})^{N^{(1)}_1}...(\partial^q\phi^{(1)})^{N_q^{(1)}}
\nonumber \\
\times
...
\times
(\phi^{(r)})^{N^{(r)}_1}...(\partial^q\phi^{(r)})^{N_q^{(r)}}
\end{eqnarray}

For simplicity, let us start from the most elementary nontrivial case
$q=r=2$, relevant to the $W^{(3)}$ IVO, whose emergence was already 
observed in the matrix model approach \cite{CRZ_2015, CR_2015}.
We shall look for the free field realization 
of this block in the form:
\begin{eqnarray}
I_{2|2}=
\sum_{N_1,N_2,N_3=0;P_1,P_2,P_3=0}^\infty\lambda_{N_1N_2N_3|P_1P_2P_3}
:\phi_1^{N_1}(\partial\phi_1)^{N_2}(\partial^2\phi_1)^{N_3}
\nonumber \\ 
\times
\phi_2^{P_1}(\partial\phi_2)^{P_2}(\partial^2\phi_2)^{P_3}:
\label{eq:I_22}
\end{eqnarray} 
with the stress-energy tensor 
\begin{equation}
T=\sum_{i=1}^2(-{1\over2}(\partial\phi_i)^2+{1\over2}Q^i\partial^2\phi_i)
\end{equation}
and the $W_3$-current
\begin{eqnarray}
j_w=\sum_{i,j,k=1}^2(\nu^i\partial^3\phi_i+\nu^{ij}\partial^2\phi_i
\partial\phi_j+\nu_{ijk}\partial\phi_i\partial\phi_j\partial\phi_k)
\label{eq:W3-current}
\end{eqnarray}
where the $\nu$-coefficients will be determined below
from the condition that $j_w$ is a dimension $3$ primary field.

The generating function $F(x_1,x_2,x_3|y_1,y_2,y_3)$
for $I_{2|2}$ is thus the function of 6 variables that is to be determined
from 3 Virasoro constraints and 3 $W^{(3)}$ constraints.
We start from the Virasoro constraints first.
As in the case of a single field, $I_{2|2}$
is the eigenvalue of $L_{2},L_{3}$ and $L_{4}$  and , since it does not
contain higher than second derivatives of the Toda fields, it is by 
construction annihilated by all higher $L_n$'s.
As before, consider the $L_4$-eigenvalue problem first.
As before, by simple straightforward calculation
the eigenvalue problem leads to the recursion
 
The constraint $[ L_4,  I_{2|2} ] = \rho_4  I_{2|2}$ results in the relation
\begin{align}
 -4 & \sum_{N_1,N_2=0,N_3=2;P_1,P_2=0,P_3=2}^\infty
(N_3(N_3-1)+P_3(P_3-1)) ~
\lambda_{N_1N_2N_3|P_1P_2P_3}
\nonumber \\
&\qquad\qquad\qquad \qquad~~~~~~~~~~~~~~~~~~~~~~
\times
:\phi_1^{N_1}(\partial\phi_1)^{N_2}(\partial^2\phi_1)^{N_3-2}
\phi_2^{P_1}(\partial\phi_2)^{P_2}(\partial^2\phi_2)^{P_3-2}:
\nonumber \\
&~~=
 \rho_4\sum_{N_1,N_2,N_3=0;P_1,P_2,P_3=0}\lambda_{N_1N_2N_3|P_1P_2P_3}
%\nonumber \\ &~~~ \times
:\phi_1^{N_1}(\partial\phi_1)^{N_2}(\partial^2\phi_1)^{N_3}
\phi_2^{P_1}(\partial\phi_2)^{P_2}(\partial^2\phi_2)^{P_3}:
\end{align}
which is equivalent to the second order PDE for the generating function:
\begin{eqnarray}
(\partial_{x_3}^2+\partial_{y_3}^2)F(x_1,x_2,x_3|y_1,y_2,y_3)
=-{{\rho_4}\over{4}}F(x_1,x_2,x_3|y_1,y_2,y_3) \,.
\label{eq:PDE_4}
\end{eqnarray}
The general solution is given as 
\begin{eqnarray}
F(x_1,x_2,x_3|y_1,y_2,y_3)
=e^{i\alpha{x_3}+i\beta{y_3}}F^{(2)}(x_1,x_2|y_1,y_2)
\end{eqnarray}
with $\alpha$ and $\beta$ coefficients satisfying
\begin{eqnarray}
\alpha^2+\beta^2={{\rho_4}\over4}
\label{eq:alpha-beta}
\end{eqnarray}
Similarly, the second eigenvalue problem, $[L_3, I_{2|2}] =\rho_3 I_{2|2}$,
leads to the  second PDE 
\begin{eqnarray}
(\partial_{x_2}\partial_{x_3}+\partial_{y_2}\partial_{y_3})F=-{{\rho_3}\over{2}}F\,.
\label{eq:PDE_3}
\end{eqnarray}
Finally, the third eigenvalue constraint, 
$[ L_2, I_{2|2} ] =\rho_2 I_{2|2}$
leads to the third PDE on $F$:
\begin{eqnarray}
2(\partial_{x_1}\partial_{x_3}+\partial_{y_1}\partial_{y_3})F
+(\partial^2_{x_2}+\partial^2_{y_2})F-(Q_1\partial_{x_3}+Q_2\partial_{y_3})F=-\rho_2F\,.
\label{eq:PDE_2}
\end{eqnarray}

The general solution of the three PDE's 
(\ref{eq:PDE_4}, \ref{eq:PDE_3}, \ref{eq:PDE_2}) 
is given in terms of the generating function
\begin{align}
F(x_1,x_2,x_3|y_1,y_2,y_3) 
=\exp{\lbrace} 
{{i\over4}\lbrack\rho_2-iQ_1\alpha-iQ_2\beta-{{({{{\rho_3}\over4}}+\lambda)^2}\over{\alpha^2}}-
{{({{{\rho_3}\over4}}-\lambda)^2}\over{\beta^2}}+{{\xi}\over{\alpha}}\rbrack{x_1}}
\nonumber \\
+{i\over4}{\lbrack\rho_2-iQ_1\alpha-iQ_2\beta-{{({{{\rho_3}\over4}}+\lambda)^2}\over{\alpha^2}}-                                                              {{({{{\rho_3}\over4}}-\lambda)^2}\over{\beta^2}}-{{\xi}\over{\beta}}
\rbrack{y_1}}
\nonumber \\
+
{i{\lbrack{{({{{\rho_3}\over4}}+\lambda)x_2}\over{\alpha}}
+{{({{\rho_3}\over4}-\lambda)y_2}\over{\beta}}+\alpha{x_3}+\beta{y_3}\rbrack}
}\rbrace
\label{eq:F-form}
\end{align}
and, accordingly,  IVO is given as 
\begin{eqnarray}
I_{2|2}=:e^{{i\over4}\lbrace\rho_2-iQ_1\alpha-iQ_2\beta-{{({{{\rho_3}\over4}}+\lambda)^2}\over{\alpha^2}}-
{{({{{\rho_3}\over4}}-\lambda)^2}\over{\beta^2}}+{{\xi}\over{\alpha}}\rbrace{\phi_1}+
i{{{({{{\rho_3}\over4}}+\lambda)\partial\phi_1}\over{\alpha}}}+i\alpha\partial^2\phi_1
}:
\nonumber \\
\times
:e^{{i\over4}\lbrace\rho_2-iQ_1\alpha-iQ_2\beta-{{({{{\rho_3}\over4}}+\lambda)^2}\over{\alpha^2}}-
{{({{{\rho_3}\over4}}-\lambda)^2}\over{\beta^2}}-{{\xi}\over{\beta}}\rbrace{\phi_2}+
i{{{({{{\rho_3}\over4}}-\lambda)\partial\phi_2}\over{\beta}}}+i\beta\partial^2\phi_2
}:
\end{eqnarray}
where 3 constants: $\lambda,\xi$ and one of $\alpha$ or $\beta$ 
(related by \eqref{eq:alpha-beta})
are not yet fixed  and must be determined from the remaining $W_3$-current constraints.

To apply the W constraint, we need to fix the coefficients  in $j_w$  current  
\eqref{eq:W3-current} first. 
To make $W_3$, $j_w$ the dimension 3 primary field, 
Generically, the OPE  of $T(z_1)$ with $j_w(z_2)$ has the form:
\begin{eqnarray}
T(z_1)j_w(z_2)\sim{z_{12}^{-5}}(12\nu_j{Q^j}-2\nu_j^j)
+z_{12}^{-4}\partial\phi_j(-6\nu^j-3\nu^{ij}_{i}+3Q_i\nu^{ij})
\nonumber \\
+z_{12}^{-3}\lbrace\partial^2\phi_j(-6\nu^j+\nu^{ji}Q_i)+\partial\phi_i\partial\phi_j
(-2\nu^{ij}+3\nu^{ijk}Q_k)\rbrace
\end{eqnarray}
(all the upper and lower indices are equivalent, distinguished merely for the 
convenience of the notations)
To make the $W_3$-current  dimension 3 we have four relations
\begin{eqnarray}
6\nu_j{Q^j}-\nu_j^j=0
\nonumber \\
-2\nu^j-\nu^{ij}_{i}+Q_i{\nu^{ij}}=0
\nonumber \\
-6\nu^j+\nu^{ji}Q_i=0
\nonumber \\
-2\nu^{ij}+3\nu^{ijk}Q_k=0\,.
\label{eq:nu-relation}
\end{eqnarray}
Note that $\nu_{ijk}$ is symmetric by construction;
a priory $\nu_{ij}$ is not necessarily symmetric, however, the last equation
in \eqref{eq:nu-relation} imposes the symmetry condition on $\nu_{ij}$.
The system  \eqref{eq:nu-relation} is thus consistent, 
being the system of 8 linear equations
for 9 variables (an extra variable corresponds to the overall normalization
of $j_w$, that is fixed by the normalization of $W^{(3)}$-algebra).
The $j_w$ current \eqref{eq:W3-current} is thus completely fixed
 by  \eqref{eq:nu-relation}.

The final step to construct the rank 2 IVO with $W^{(3)}$-symmetry
is to solve the eigenvalue problems for $I_{2|2}$ with respect to
$j_w$ modes: $j_w(z)=\sum_{n}z^{-n-3}W_n^{(3)}$.
Namely, $I_{2|2}$ must be the simultaneous eigenvector of $W_k^{(3)}$ with $k=4, 5, 6$
and annihilated by higher modes. As in the Virasoro case, the 
annihilation constraint is automatically ensured by the manifest form of
the ansatz \eqref{eq:I_22}. The W-constraints on $I_{2|2}$ lead
to extra 3 linear partial differential equations of the third order on the 
generating function $F$,
allowing to fix the remaining unknown constants in \eqref{eq:F-form}. 
Namely, applying \eqref{eq:W3-current} to \eqref{eq:I_22} and proceeding precisely as  explained above,
we  obtain the system of 3 extra differential equations on $F$.
For the eigenvalue problem 
\begin{equation}
[W_6^{(3)}, I_{2|2} ]=\omega_6 {I_{2|2}};
\end{equation}
we have 
\begin{eqnarray}
(3\nu_{111}\partial_{x_3}^3+2\nu_{112}\partial_{x^3}^2\partial_{y_3}+2\nu_{122}\partial_{x_3}\partial_{y_3}^2
%\nonumber \\
+3
\nu_{222}\partial_{y_3}^3+{{\omega_6}\over8})F(x_1,x_2,x_3|y_1,y_2,y_3)
=0 \,.
\end{eqnarray}
For the eigenvalue problem 
\begin{equation}
[W_5^{(3)}, I_{2|2} ]=\omega_5 {I_{2|2}};
\end{equation}
we have 
\begin{align}
\lbrace
\nu_{111}(6\partial^2_{x_3}\partial_{x_1}+3\partial^2_{x_2}\partial_{x_3})
%\nonumber \\
&+\nu_{112}(2\partial_{x_3}\partial_{y_3}\partial_{x_1}+\partial^2_{x_2}\partial_{y_3}
+\partial_{x_2}\partial_{x_3}\partial_{y_2}+2\partial^2_{x_3}\partial_{y_1})
\nonumber \\
&+\nu_{122}(2\partial_{x_3}\partial_{y_3}\partial_{y_1}+\partial^2_{y_2}\partial_{x_3}
+\partial_{y_2}\partial_{y_3}\partial_{x_2}+2\partial^2_{y_3}\partial_{x_1})
\nonumber \\ 
&+\nu_{222}(6\partial^2_{y_3}\partial_{y_1}+3\partial^2_{y_2}\partial_{y_3})
+{{\omega_5}\over4}\rbrace{F}(x_1,x_2,x_3|y_1,y_2,y_3)=0\,.
\end{align}
And finally, for the eigenvalue problem 
\begin{equation}
[W_4^{(3)}, I_{2|2} ] =\omega_4 {I_{2|2}};
\end{equation}
we have 
\begin{eqnarray}
\lbrace
\nu_{111}(6\partial_{x_3}^2\partial_{x_1}+3\partial_{x_2}^2\partial_{x_3})
+\nu_{112}(2\partial_{x_1}\partial_{x_3}\partial_{y_3}
+
\partial_{x_2}\partial_{x_3}\partial_{y_2}+2\partial_{x_3}^2\partial_{y_1})
\nonumber \\
+\nu_{122}
(2\partial_{x_3}\partial_{y_1}\partial_{y_3}
+
\partial_{x_3}\partial^2_{y_2}
+2
\partial_{x_1}\partial^2_{y_3})
+
\nu_{222}(6\partial_{y_3}^2\partial_{y_1}+3\partial_{y_2}^2\partial_{y_3})
\nonumber \\
+3\nu_{11}\partial_{x_3}^2+6\nu_{12}\partial_{x_3}\partial_{y_3}
+3\nu_{22}\partial^2_{y_3}+{{\omega_4}\over4}\rbrace
F(x_1,x_2,x_3|y_1,y_2,y_3)=0\,.
\end{eqnarray}

From the 3 PDE with the form $F$ in \eqref{eq:F-form} give the following algebraic constraints on the
remaining constants:
\begin{eqnarray}
3\nu_{111}\alpha^3+3\nu_{222}\beta^3+\nu_{112}\alpha^2\beta
+\nu_{122}\alpha\beta^2+{{i\omega_6}\over{16}}=0
\end{eqnarray} 
\begin{eqnarray}
3\nu_{111}\alpha({{\rho_3}\over{4}}+\lambda)
+
3\nu_{222}\beta({{\rho_3}\over{4}}-\lambda)
+\nu_{112}({{\alpha^2}\over{\beta}}({{\rho_3}\over{4}}-\lambda)
+\beta({{\rho_3}\over{4}}+\lambda))
\nonumber \\
+\nu_{122}({{\beta^2}\over{\alpha}}({{\rho_3}\over{4}}+\lambda)
+\alpha({{\rho_3}\over{4}}-\lambda))
+{{i\omega_5}\over8}=0
\end{eqnarray}
\begin{eqnarray}
\nu_{111}(
{3\over2}\alpha^2(\rho_2-iQ_1\alpha-iQ_2\beta
-{1\over{\alpha^2}}({{\rho_3}\over{4}}+\lambda)^2
\nonumber \\
+{1\over{\beta^2}}({{\rho_3}\over{4}}-\lambda)^2)
+3\beta({{\rho_3}\over{4}}+\lambda)+{{\xi}\over{\alpha}})
\nonumber \\
+
\nu_{222}(
{3\over2}\beta^2(\rho_2-iQ_1\alpha-iQ_2\beta
-{1\over{\alpha^2}}({{\rho_3}\over{4}}+\lambda)^2
\nonumber \\
+ 
{1\over{\beta^2}}({{\rho_3}\over{4}}-\lambda)^2)
+3\alpha({{\rho_3}\over{4}}+\lambda)-{{\xi}\over{\beta}})
\nonumber \\
+
\nu_{112}({1\over2}\alpha\beta(\rho_2-iQ_1\alpha
-iQ_2\beta-{1\over{\alpha^2}}({{\rho_3}\over4}+\lambda)^2
+{1\over{\beta^2}}({{\rho_3}\over4}-\lambda)^2+{{\xi}\over{\alpha}})+
\nonumber \\
+ {{\beta}\over{\alpha^2}}
({{\rho_3}\over4}+\lambda)^2
+{1\over{\beta}}({{\rho_3}\over4}+\lambda)({{\rho_3}\over4}-\lambda)
\nonumber \\
+{{\alpha^2}\over2}
(\rho_2-iQ_1\alpha-iQ_2\beta
-{1\over{\alpha^2}}({{\rho_3}\over{4}}+\lambda)^2
+
{1\over{\beta^2}}({{\rho_3}\over{4}}-\lambda)^2-{{\xi}\over{\beta}})
\nonumber \\
\nu_{122}({1\over2}
\alpha\beta(\rho_2-iQ_1\alpha-iQ_2\beta-{1\over{\alpha^2}}
({{\rho_3}\over4}+\lambda)^2
+{1\over{\beta^2}}({{\rho_3}\over4}-\lambda)^2-{{\xi}\over{\beta}})
\nonumber \\
+ {{\alpha}\over{\beta^2}}
({{\rho_3}\over4}-\lambda)^2
+{1\over{\alpha}}({{\rho_3}\over4}+\lambda)({{\rho_3}\over4}-\lambda)
\nonumber \\
+{{\beta^2}\over2}
(\rho_2-iQ_1\alpha-iQ_2\beta
-{1\over{\alpha^2}}({{\rho_3}\over{4}}+\lambda)^2+
{1\over{\beta^2}}({{\rho_3}\over{4}}-\lambda)^2+{{\xi}\over{\alpha}}))
\nonumber \\
+3\nu_{11}\alpha^2+3\nu_{22}\beta^2+6\nu_{12}\alpha\beta
+{{i\omega_4}\over4}=0
\end{eqnarray}
This system of cubic algebraic equations fixes the remaining coefficients and fully defines the 
Virasoro and $W_3$ irregular vertex operator.

\section{Conclusion} 

In this paper we have constructed an explicit form of the irregular vertex operator 
with Virasoro and  $W$-symmetry.
Given the irregular vertex operators, constructed in this work, it is straightforward  to read off 
the associate Penner type potentials
whose random matrix model 
has the Seiberg-Witten curves corresponding to the 4d gauge theories.
Although in the text we limited the explicit examples to the Virasoro cases and to the $W_3$-case of rank 2, it is not difficult to see the pattern for the general $W^{(N)}$ with arbitrary rank $q$.
The vertex operators  would generally contain $N-1$ Toda fields and involve the derivatives of  orders up  to $q$:
\begin{eqnarray}
I_{N|q}=:e^{\sum_{a=1}^{N-1}\sum_{k=0}^q\alpha_{a|k}\partial^k\phi^{(a)}}:
\label{eq:IVO_Ws-q}
\end{eqnarray}
This IVO's again generates the simultaneous eigenstate of $L_n$
for $q\leq{n}\leq{2q}$ with eigenvalues $\rho_n$ (annihilated by the higher $L_n$'s)
and of the expansion modes $W_n^{(s)}$ for $(s-1)q\leq{n}\leq{sq}$
with eigenvalues $\lambda_n^{(s)}$ (annihilated by higher $W_n^{(s)}$).
The $W$ current with the integer spin  $3\leq{s}\leq{N}$ has the form 
\begin{eqnarray}
j_w^{(s)}(z)\equiv\sum_{n}{{W_n^{(s)}}\over{z^{n+s}}}
=\sum_{r=1}^N\sum_{s|p_1...p_r}\sum_{\lbrace{a_1,...,a_r\rbrace}}\nu^{(s)}_{a_1...a_r|p_1...p_r}
\partial^{p_1}\phi^{(a_1)}...\partial^{p_r}\phi^{(a_r)}
\end{eqnarray}
where the sum is taken over the ordered partitions of 
$s=p_1+...+p_r; 1\leq{p_1}...\leq{p_r}$ with the lengths $1\leq{r}\leq{N}$
and $1\leq{a_1}\leq{a_2}....\leq{a_r}\leq{N-1}$.
The coefficients $\nu^{(s)}_{a_1...a_r|p_1...p_r}$ are determined by $N-2$ systems of
linear algebraic equations (one per each $s$ ) stemming from the primary field constraints for each $j_w^{(s)}$. Once the $\nu$-coefficients are fixed,
the $\alpha_{a|k}$ coefficients are related to 
 $\nu^{(s)}_{a_1...a_r|p_1...p_r}$  and the eigenvalues $\rho_n$ and $\lambda_n^{(s)}$
by the system  $(N-1)(q+1)$ algebraic (nonlinear) equations, exactly matching the number   of the coefficients.

These algebraic constraints altogether (for $\nu$ and for $\alpha$) fully determine
the  the $W_N$ irregular blocks related to the degree $N$ Seiberg-Witten curves in 
Argyres-Douglas theories.
Given the coefficients in the irregular vertex operators, it is straightforward to establish their relation to eigenvalues of Virasoro generators and  $W$ generators.
For simplicity, we shall demonstrate it for the  $W^{(3)}$ irregular vertex operator of an arbitrary rank.
However, the computation below is straightforward to establish for the arbitrary $n$ case.
Let's consider the irregular vertex operator  \eqref{eq:IVO_Ws-q} for the $W^{(3)}$-case 
and expand it in series of $\phi$ and its derivatives:
\begin{eqnarray}
I_{3|q}=:e^{\sum_{a=1}^{2}\sum_{k=0}^q\alpha_{a|k}\partial^k\phi^{(a)}}
=\prod_{a=1}^2\prod_{k_a=0}^q\sum_{N_{a|k_a}=0}^\infty
{{(\alpha_{a|k_a}\partial^{k_a}\phi_a)^{N_{a|k_a}}}\over{N_{a|k_a}!}}
\label{eq:IVO_W3-q}
\end{eqnarray}
Applying the stress-energy tensor to \eqref{eq:IVO_W3-q} and re-exponentiating we obtain
for $[L_r, I_{3|q}]=\rho_r I_{3|q}$ with $q\leq{r}\leq{2q}$ where 
\begin{eqnarray} 
\rho_r=  -\sum_{a=1}^2\sum_{p_a+q_a=r;0\leq{p_a,q_a}\leq{q}}{(p_a)!(q_a)!}\alpha_{a|p_a}\alpha_{a|q_a}
+{Q\over2}(p_a+1)!\alpha_{a|p_a}\delta_{p_a|q}
\end{eqnarray}
Finally, applying $[W_r^{(3)}, I_{3|q}]=\omega_r I_{3|q}$ we have  
\begin{align} 
\omega_r=-
\sum_{i,j,k=1}^2\lbrace\sum_{m_i+p_j+q_k=r;0\leq{m_i,p_j,q_k}\leq{q}}
\sigma^{ijk}\nu^{ijk}
m_i!p_j!q_k!\alpha_{i|m_i}\alpha_{j|p_j}\alpha_{k|q_k}
\nonumber \\
+
\sum_{m_i+p_j=r;0\leq{m_i,p_j}\leq{q-1}}
m_i!(p_j+1)!\alpha_{i|m_i}\alpha_{j|p_j}
%\nonumber \\
+
(r+2)!\nu^i\alpha_{i|r}\rbrace
\end{align}
where $\sigma^{ijk}$ is the symmetric factor, symmetric in the $i,j,k$ and 
 $\sigma^{111}=\sigma^{222}=3!$, $\sigma^{112}=\sigma^{122}=2!$. 
The relations (4.4) and (4.5) reproduce those obtained earlier in
\cite{lefloch} using a different approach, by direct application of the 
operator product expansion  to the colliding limit of regular vertex operators. 

It is straightforward to check that the irregular states created by the irregular vertex 
operators are coherent, i.e. are the eigenstates of the spin 1 conserving current
$\partial\phi$ (note that, just as spin 2 conserving  current $T(z)$, conserving spin 1 is not
a primary field if $Q\neq{0}$).
Indeed, expanding the general irregular operator \eqref{eq:IVO_Ws-q}  in series similarly to \eqref{eq:IVO_W3-q} 
and re-exponentiating one it is easy to verify the OPE
\begin{eqnarray}
\partial\phi^{(b)}(z) I_{N|q}(w)
= - \sum_{k=0}^q{{\alpha_{b|k}k!}\over{(z-w)^{k}}}I_{N|q}(w)+ regular
\end{eqnarray}
from which the coherent state property follows.

The exponents for the irregular vertices of the type  \eqref{eq:IVO_W3-q}, 
whose explicit examples have been constructed in our work, can of course 
be expanded in powers of the  derivatives of  $\phi^{(a)}$, leading to combinations
of these derivatives acting on $regular$  vertex operators
$e^{\sum_{a}\alpha_{a|0}\phi^{(a)}}$ in Toda theories.
These terms can  be classified according to total conformal dimensions $h$ 
carried by the derivatives acting on the regular vertex.
Each dimension $h$'s contribution to the expansion can
be cast as some combination of products
of the negative Virasoro and $W$-current modes 
${\sim}L_{-h_1}...L_{-h_p}W^{(s_1)}_{-h_{p+1}}W^{(s_q)}_{-h_{p+q}}$
acting on the regular vertex.
where $h_k;k=1...p+q$ are the elements of the length $p+q$ partitions
of $h$.

This generalizes the expansion of the irregular states in terms of the 
Virasoro descendants of the primaries created by regular vertex operators,
discussed in \cite{KMST_2013, CRZ_2015, CR_2015} to $W^{(N)}$-case.
Note that, in this descendent expansion approach,
the expansion coefficients were not completely fixed even in the rank 2 Virasoro case 
from the eigenvalue constraint. One needs further consistency conditions
with the lower Virasoro mode \cite{CRZ_2014}. 
As seen in this free field approach, the expansion coefficients for the irregular vertex operators 
should be determined completely without resorting to other conditions. 
The difficulty simply is related with the fact that if the descendent decomposition 
has more varaibles than the number of eigenvalue constraints.  

The irregular vertex operators and the irregular blocks, studied in this paper, appear to be 
quite fascinating objects by themselves, and  may be of interest  far beyond
AGT conjecture and Liouville/Toda theories.
First of all, from the AdS/CFT point of view it seems plausible that the irregular blocks may be string-theoretic
duals of some important classes of local composite operators on the 
gauge/CFT side, e.g. such as $\sim{T_{\mu_1\nu_1}...T_{\mu_n\nu_n}}$.
On the other hand, 
the operators of this sort must correspond to higher spin modes in $AdS$ with mixed symmetries.
 As $T_{\mu\nu}$ is the CFT dual of the graviton vertex operator
\cite{GKP_1998}, the operators like $\sim{T^n}$ can be understood as the colliding limit
of $n$ gravitons, i.e. a rank $n$ irregular block, generalized to string theory.
Being  non-primaries, these objects are of course not in the BRST cohomology and 
therefore are essentially off-shell. On the other hand, they constitute a subclass
of operators which is far richer than the subspace of primaries, but appear to have very nice
and controllable behaviour under global conformal transformations.
As such,  they may play an important role in string field theory (SFT), being
crucial elements for finding new classes of analytic solutions. 
Given that SFT is currently  our best hope to advance towards background independent formulation
of string theory,  and that analytic solutions constitute a crucial ingredient 
in such a formulation, one can anticipate that the irregular blocks may be
of  importance and interest in describing various nonperturbative backgrounds
in string theory (such as collective higher spin vacuum states).
Ultimately, the deeper undestanding of the irregular blocks may be important step 
towards the understanding  the interplays between two-dimensional and four-dimensional theories
which at the moment still largely retain the status of conjectures.

\subsection*{Acknowledgements}
This work is partially supported by the National Research Foundation of Korea(NRF) grant funded by the 
Korea government(MSIP) (NRF-2014R1A2A2A01004951) 
and by the National Natural Science Foundation of China under grant 11575119. 
We would like to thank 
Bruno Le Floch for useful remarks and comments, as well as for pointing out
some typos in the initial version of the manuscript. We also thank B. Le Floch and H. Nagoya
for pointing us out the references containing
discussions and results related to those investigated in this work.

%================================%

\end{document}